\documentclass[twocolumn,showpacs,amssymb,prd]{revtex4}

\usepackage{graphicx}% Include figure files
\usepackage{dcolumn}% Align table columns on decimal point
\usepackage{bm}% bold math

\def\beq{\begin{equation}}
\def\enq{\end{equation}}
\def\ba{\begin{eqnarray}}
\def\ea{\end{eqnarray}}
\def\Mesz{M\'esz\'aros~}
\def\<{<\!\!}
\def\>{\!\!>}
\def\ra{\rightarrow}
\def\eps{\epsilon}

\def\msun{M_\odot}

\begin{document}
\input{epsf}

\title{Neutrino signatures of the supernova - gamma ray burst
relationship}

\author{Soebur Razzaque,$^1$ Peter M\'esz\'aros$^1$ and Eli
Waxman$^2$}

\affiliation{$^1$Dpt Astronomy \& Astrophysics, Dpt Physics,
 Pennsylvania State Univ., University Park, PA 16802, USA \\
 $^2$Department of Condensed Matter Physics, Weizmann Institute of
 Science, Rehovot 76100, Israel}

\begin{abstract}
We calculate the TeV-PeV neutrino fluxes of gamma-ray bursts
associated with supernovae, based on the observed association between
GRB 030329 and supernova SN 2003dh. The neutrino spectral flux
distributions can test for possible delays between the supernova and
the gamma-ray burst events down to much shorter timescales than what
can be resolved with photons. As an illustrative example, we calculate
the probability of neutrino induced muon and electron cascade events
in a km scale under-ice detector at the South Pole, from the GRB
030329.  Our calculations demonstrate that km scale neutrino
telescopes are expected to detect signals that will allow to constrain
supernova-GRB models.
\end{abstract}

\date{\today}

\maketitle

\section{Introduction}

Gamma-ray bursts (GRB) have recently been shown to be associated with
supernova (SN) events. This is based, most notably, on the burst GRB
030329, in whose optical afterglow the supernova (labelled SN 2003dh)
photometric and spectral signatures show up 9.6 days \cite{stanek03}
after the $\gamma$-ray trigger. This late brightening of the supernova
is caused by the large initial opacity to optical photons of the
envelope of the exploding supernova. The collapse of the stellar core
is thought to lead, for progenitor stellar masses in excess of 30
$\msun$, to a blackhole responsible for the GRB (e.g. \cite{mesz02}
for a review), while for less massive progenitors it leads to a
rotating neutron star, typically detectable as a pulsar. A question of
great interest is whether the supernova explosion of a progenitor in
excess of $\sim 30\msun$ can lead to a black hole promptly (on a
timescale comparable to the core collapse free-fall time), or else
after some delay, during which an initial pulsar accretes additional
gas which falls back onto it from the envelope \cite{macfadyen01}. In
the extreme version of this scenario the collapse to a black hole is
delayed by weeks to months \cite{vie98}. Optical observations cannot
test the time off-set which characterizes this scenario to an accuracy
better than a few days \cite{stanek03, matheson03, SN2003dh}, due to
observational difficulties and uncertainties in modeling the diffusive
photon transport in the dynamically evolving envelope. Neutrinos,
however, are not subject to the very long diffusion times of photons,
due to their much lower opacities, and should provide a much tighter
time-tracking of the collapse.  While thermal (MeV) neutrinos produced
in the collapse are presently undetectable except from the very
nearest galaxies, ultra-high energy ($\gtrsim$ TeV) neutrinos from
GRBs are expected to be detectable from cosmological distances with
kilometer scale Cherenkov detectors \cite{icecube, rice, anita,
antares} or giant air-shower detector \cite{auger}. Here we calculate
the $\gtrsim$ TeV neutrino signatures associated with the collapse of
massive stellar progenitors of GRB, both in the case where a supernova
envelope is ejected simultaneously with, or sometime before, the GRB
event. We calculate the resulting muon and electron cascade events in
kilometer scale sub-ice detctors, and discuss the characteristic
signatures associated with various time off-sets between the GRB and
the SN.

\section{Observations and model}
\label{sec:model}

The GRB 030329 is located at the coordinates $\alpha = 10^h 44^m
50.03^s$, $\delta = +21^{\circ} 31' 18.1''$ (J2000)
\cite{stanek03}. The distance of the burst at redshift $z \approx
0.17$ is $D \approx 10^{27.3} D_{27.3}$ cm for an $\Omega_m = 0.3$,
$\Omega_\Lambda=0.7$ and $H_o = 75$ km s$^{-1}$ Mpc$^{-1}$ cosmology.
The observed GRB duration is $\Delta t \sim 30 \Delta t_{30}$ s with
$10^{-4}$ erg/cm$^2$ fluence in the 30-400 keV range.  The peak flux
is $7 \times 10^{-6}$ erg cm$^{-2}$ s$^{-1}$ with 1.2 s duration. The
isotropic $\gamma$-ray luminosity $L_{\gamma}^{\rm iso} \sim
10^{51}L_{\gamma,51}$ erg/s is then approximately the same as the
typical value for long bursts.

We take the progenitor star to have an original mass of $\sim 30-40
M_{\odot}$, which subsequently lost its H envelope leaving behind a He
core of $\sim 14 M_{\odot}$ in its presupernova phase. The He core has
initially a smaller radius of $\sim 1.2 R_{\odot}$, for a solar
metallicity star, but grows larger later in its He burning stage.  The
density profile of the star can be roughly estimated as $\rho = \rho_*
(R_*/r -1)^n$, where $\rho_* = 2$ g/cm$^3$ and $R_* \sim 10^{11}$ cm
in the presupernova phase \cite{wm03}. In case of $n \approx 3$, for
radiative envelope, $\rho \approx 10^{-3}$ g/cm$^3$ for $r \sim
0.9R_*$.

The supernova SN 2003dh is similar to SN 1998bw, with a similar
explosion energy estimated as $E_{sn} = (2 - 5)\times 10^{52}$ erg,
and an ejected supernova remnant (SNR) shell velocity of $\approx
0.1c$ \cite{matheson03}. We consider here two scenarios: first, the SN
takes place 0.1-8 days prior to the (electromagnetically detected) GRB
event; and second, both the SN and the GRB occur simultaneously. The
neutrino fluxes from these two scenarios differ significantly, as
discussed in the following section.

\section{Neutrino flux components}
\label{sec:nu-flux}

High energy neutrinos are created from photomeson ($p\gamma$) and
proton-proton ($pp$) interactions by the shock accelerated protons in
the GRB shocks. In the $\gamma$-ray prompt phase of the GRB, neutrinos
are produced through $p\gamma$ interactions \cite{wb-burst}. A
precursor component arises, due to $pp$ and $p\gamma$ interactions,
when the GRB jet is burrowing it's way through the star \cite{mw01,
rmw03b}. If an SNR shell is ejected days before the GRB event, an
extra component is added to the prompt phase neutrino flux becasue of
$pp$ and $p\gamma$ interactions in the shell \cite{rmw03a,
gg03}. Afterglow reverse shocks also generate high energy neutrinos
through $p\gamma$ interactions \cite{wb00, Dai00}. We consider each of
these flux components, in the framework of astrophysical models
described in Sec. \ref{sec:model}, below.

\subsection{Burst}
\label{subsec:burst}

The Thomson depth of the GRB internal shocks at a radius $r_{\rm sh}$
is $\tau_{\rm Th} \approx \sigma_{\rm Th} L_{\gamma}^{\rm iso} \delta
t/(4\pi r_{\rm sh}^2 \Gamma m_p c^2)$, where $\delta t \sim 10^{-3}
\delta t_{-3}$ s is the variability time and $\Gamma \approx 300
\Gamma_{300}$ is the bulk Lorentz factor. Solving for $r_{\rm sh} =
r_{\rm thin}$ at $\tau_{\rm Th}=1$, we get the shock radius at which
the GRB jet becomes optically thin as
\ba 
r_{\rm thin} \approx 10^{12} \left( \frac{L_{\gamma,51}\delta
t_{-3}}{\Gamma_{300}} \right)^{1/2}~{\rm cm}.
\label{jetradius-thin} 
\ea
The optical depth for $p\gamma \ra \Delta$ at threshold with observed
$\eps_{\gamma}^{\rm burst} \approx 0.5$ MeV peak synchrotron photons
is
\ba 
\tau_{p\gamma} \sim 10 \left( \frac{\sigma_{p\gamma}}{\sigma_{\rm Th}}
\right) \left( \frac{\eps_{\gamma}^{\rm burst}}{0.5~\rm{MeV}}
\right)^{-1} \Gamma_{300}.
\label{photo-opt} 
\ea
Correspondingly, the proton to pion conversion efficiency in the
internal shocks is $f_{\pi}^{\rm burst} \approx 1f_{\pi,0}^{\rm
burst}$. The neutrino break energy for $p\gamma$ interactions with
peak synchrotron photons is
\ba 
E_{\nu,b}^{\rm burst} \approx 2\times 10^6 \left(
\frac{\eps_{\gamma}^{\rm burst}} {0.5~\rm{MeV}} \right)^{-1}
\Gamma_{300}^2~\rm{GeV;}
\label{burst-nubreak}
\ea
assuming 5\% of the proton energy goes to a neutrino in each
interaction at the $\Delta$-resonance ($E_p \eps_{\gamma}^{\rm burst}
= 0.2$ GeV$^2$ in the comoving frame). Neutrinos below this energy
interact with synchrotron photons above $\eps_{\gamma}^{\rm burst}
\approx 0.5$ MeV which follow a typical power law distribution
$dN_{\gamma}/dE_{\gamma} \propto E_{\gamma}^{-2}$. At ultra-high
energies, neutrino production is suppressed due to synchrotron losses
of secondary pions and muons produced from $p\gamma$ interactions. The
maximum pion synchrotron break energy, from equipartition magnetic
field with equipartition fraction $\xi_B = 0.01 \xi_{B,-2}$, is
\ba 
E_{\pi, sb}^{\rm burst} \approx 3.8\times 10^8 \left(
\frac{L_{\gamma,51} \Gamma_{300} \delta t_{-3}}{\xi_{B,-2}}
\right)^{1/2} ~\mbox{GeV}.
\label{pi-syncloss}
\ea
Assuming the luminosity of internal shock-accelerated protons in the
GRB is $L_p \sim \kappa L_{\gamma,51}$, the power-law distribution of
protons is
\ba
\frac{d^2 N_p}{dE_p dt} \approx 6.2\times 10^{53}~\frac{
\xi_p \kappa L_{\gamma,51}} {E_{p}^2}~{\rm GeV}^{-1}~{\rm s}^{-1}. 
\ea
Here $\xi_p \lesssim 1$ is the fraction of protons accelerated and
$\kappa \gtrsim 1$. The corresponding neutrino spectrum at earth
[$\Phi_{\nu} = d^2N_{\nu}/dE_{\nu} dt = (4\pi D^2)^{-1} d^2N_p/dE_p
dt$] is
\ba
E_{\nu}^2 \Phi_{\nu} &=& 3.1\times 10^{-3} \frac{f_{\pi,0}^{\rm burst}
\xi_p \kappa L_{\gamma,51}} {D_{27.3}^{2}} \nonumber \\ && \times
\cases{ (E_{\nu}/E_{\nu,b}^{\rm burst}); \;\; E_{\nu} < E_{\nu,b}^{\rm
burst} \cr 1; \;\; E_{\nu,sb}^{\rm burst} > E_{\nu} > E_{\nu,b}^{\rm
burst} \cr (E_{\nu}/E_{\nu,sb}^{\rm burst})^{-1}; \;\; E_{\nu} >
E_{\nu,sb}^{\rm burst} } \nonumber \\ && \times \;
\rm{GeV~cm}^{-2}~\rm{s}^{-1},
\label{burst-nuflux}
\ea 
where $E_{\nu,sb}^{\rm burst} \approx 9.4\times 10^7$ GeV is the
maximum synchrotron energy for neutrinos from Eq. (\ref{pi-syncloss}).

\subsection{Afterglow}
\label{subsec:afterglow}

The GRB afterglow arises as the jet fireball ejecta runs into the
ambient medium [the inter-stellar medium (ISM) or a pre-ejected
stellar wind], driving a blast wave ahead into it and a reverse shock
back into the GRB jet ejecta. This (external) reverse shock takes
place well beyond the internal shocks, at a radius $r_e \sim
4\Gamma_e^2 c \Delta t \sim 10^{17.3} \Gamma_{250}^2 \Delta t_{30}$ cm
\cite{wb00}.  Here $\Gamma_{e} \approx 250 \Gamma_{250}$ is the bulk
Lorentz factor of the ejecta after the partial energy loss incurred in
the internal shocks.  Neutrinos are produced in the external reverse
shock due to $p\gamma$ interactions of internal shock accelerated
protons predominantly with synchrotron soft x-ray photons produced by
the reverse shock. The characteristic photon energy in the observer
frame for the ISM case is \cite{wb00}
\ba
\eps_{\gamma}^{\rm glow} \approx \frac{0.3}{ \xi_{B,-2}^{3/2} n_0
L_{\gamma,51}^{1/2} \Delta t_{30}}~\rm{keV,}
\label{reverse-photE}
\ea
where $0.01 \xi_{B,-2}$ is the magnetic field equipartition fraction
from typical afterglow fits and $1 n_0$ cm$^{-3}$ is the inter-stellar
medium density. The corresponding neutrino break energy, for $p\gamma$
interactions, is
\ba
E_{\nu,b}^{\rm glow} \approx 2\times 10^{9} \left(
\frac{\eps_{\gamma}^{\rm glow}}{0.3~\rm{keV}} \right)^{-1}
\Gamma_{250}^2~\rm{GeV.}
\label{glow-nubreak}
\ea

The efficiency of pion conversion from $p\gamma$ interactions in GRB
afterglows is $f_{\pi}^{\rm glow} \approx 0.1 f_{\pi,-1}^{\rm glow}$;
smaller than in internal shocks \cite{wb00}. The afterglow neutrino
spectrum [Afterglow (ISM) case] is then
\ba
E_{\nu}^2 \Phi_{\nu} &=& 3.1 \times 10^{-3}~\frac{f_{\pi,-1}^{\rm
glow} \xi_p \kappa L_{\gamma,51}} {D_{27.3}^{2}} \nonumber \\ &&
\times \cases{ (E_{\nu}/E_{\nu,b}^{\rm glow}); \;\; E_{\nu} <
E_{\nu,b}^{\rm glow} \cr (E_{\nu}/E_{\nu,b}^{\rm glow})^{1/2}; \;\;
E_{\nu} > E_{\nu,b}^{\rm glow} } \nonumber \\ &&
\times~\rm{GeV~cm}^{-2}~\rm{s}^{-1}.
\label{glow-nuflux}
\ea
Note that ultra-high energy neutrinos from afterglows are not
suppressed due to pion synchrotron losses, because of the smaller
magnetic fields at the larger radius of the external shock (in
constrast to internal shocks).

The estimate in Eq. (\ref{glow-nuflux}) is derived under the
assumption that the jet expands into a medium with a density typical
to that of the interstellar medium, $n\simeq 1~{\rm
cm}^{-3}$. However, in the case of a massive star progenitor the jet
may be expanding into a wind, emitted by the progenitor prior to its
collapse. In this case, the density of the surrounding medium, at the
external shock radius, may be much higher than that typical to the
interstellar medium. For a wind with mass loss rate of
$10^{-5}M_\odot~{\rm yr}^{-1}$ and velocity of $v_w=10^3~{\rm km/s}$,
the wind density at the typical external shock radius would be $\simeq
10^4~{\rm cm}^{-3}$. The higher density implies a lower Lorenz factor
of the expanding plasma during the reverse shocks stage, and hence a
larger fraction of proton energy lost to pion production. Protons of
energy $E_p\gtrsim 10^{18}$~eV lose all their energy to pion
production \cite{wb00, Dai00}, and the resulting neutrino flux
[Afterglow (wind) case] is approximately given by~\cite{w02}
\ba 
E_\nu^2\Phi_\nu &\approx & 3.1\times 10^{-3}~\frac{\xi_p \kappa
L_{\gamma,51}} {D_{27.3}^{2}}~\min \left(1,~\frac{E_\nu}{10^{17}~{\rm
eV}} \right) \nonumber \\ && \times~{\rm GeV}~{\rm cm^{-2}~s^{-1}}.
\label{eq:JGRBAGw}
\ea
The neutrino flux is expected to be strongly suppressed at energies
$E_\nu>10^{19}$~eV, since protons are not expected to be accelerated
to energies $E_p\gg10^{20}$~eV.

\subsection{Supranova}
\label{subsec:supranova}

If the SN explosion resulting in a SNR shell ejection takes place
hours to few days prior to the GRB (``supranova" scenario), then one
expects an extra neutrino component, due to GRB-accelerated proton
interactions taking place in the pre-ejected SNR shell, which is in
addition to the burst and afterglow components.  Optical observations
suggest a maximum of 2-8 days delay \cite{matheson03, SN2003dh}, and
are compatible with no delay.

Assuming the SN energy $E_{\rm sn} \approx 2\times 10^{52}E_{52.3}$
erg converts to kinetic energy with the observed velocity $v_{\rm snr}
\approx 3\times 10^{9}v_{9.5}$ cm/s of the SNR shell, the mass of the
shell is $M_{\rm snr} \approx 2-3 M_{\odot}$. For our calculation, we
take $M_{\rm snr} = 2.2 m_{2.2} M_{\odot}$. The typical distance
reached by the SNR shell is $R_{\rm snr} \approx 10^{14.4} v_{9.5}
t_d$ cm in $t_d$ days.  We assume an isotropic distribution of shell
matter with $\delta = \Delta R_{\rm snr}/R_{\rm snr} =0.1 \delta_{-1}$
width for $t_d \gtrsim 1$ day. For younger SNR shells, the matter may
be taken to be approximately isotropically distributed between the
progenitor star and up to $R_{\rm snr}$.

One of the two dominant photon components trapped inside the SNR shell
consists of photons from the SN shock with black body temperature of
\cite{rmw03a}
\ba
\eps_{\gamma}^{\rm sn} \approx 5~\frac{ E_{52.3}^{1/4}
\delta_{-1}^{1/3} R_*^{1/4}} {v_{9.5} t_d}~{\rm eV}.  
\ea
Another photon component may arise, for $t_d \gtrsim 1$ day, because
of a magnetohydrodynamic (MHD) wind which impacts the inner shell
radius driving a forward shock through the shell.  With an MHD wind
luminosity of $L_{\rm mhd} \sim 10^{46}L_{\rm m46}$ erg/s from a
fast-rotating pulsar resulting from the SN, the thermalized photon
energy in the SNR shell is 
\ba
\eps_{\gamma}^{\rm m} \approx 17~\frac{L_{\rm m46}^{1/4}
\delta_{-1}^{1/4}} {v_{9.5}^{3/4} t_d^{1/2}}~{\rm eV}.  
\ea
The optical depths for $p\gamma$ interactions of these two photon
components are
\ba
\tau_{p\gamma}^{\rm sn} &\approx & 8\times 10^4 
~\frac{E_{52.3}^{3/4}
\delta_{-1}^{1/3}} {R_{*}^{1/4} v_{9.5} t_d} \nonumber \\
\tau_{p\gamma}^{\rm m} &\approx & 10^4 
~\frac{L_{\rm m46}^{3/4}
\delta_{-1}^{1/4}} {v_{9.5}^{1/2} t_d^{1/2}}.
\ea

High energy protons escaping from GRB internal shocks (with an escape
factor $\eta_p < 1$) may interact with thermalized SN and MHD photons
in the SNR shell. The corresponding neutrino threshold energies are
\ba
E_{\nu,th}^{\rm sn} &\approx & 5.7 \times 10^8 ~\frac{ v_{9.5} t_d
\Gamma_{300}} {E_{52.3}^{1/4}\delta_{-1}^{1/3} R_{*}^{1/4}}
~\hbox{GeV.}  \nonumber \\ E_{\nu,th}^{\rm m} &\approx & 1.7\times
10^8 ~\frac{v_{9.5}^{3/4}t_d^{1/2} \Gamma_{300}} {L_{\rm m46}^{1/4}
\delta_{-1}^{1/4}} ~\hbox{GeV}.
\label{snr-nuthresh}
\ea
in the observer frame. Protons below $\Delta$ production threshold
energies may undergo $pp$ interactions in the SNR shell. The mean
optical depth for such $pp$ interactions is $\< \tau_{pp} \> \approx
60~\zeta_{sh} m_{2.2}/(v_{9.5}t_d)^{2}$ for an average 60 mb total
$pp$ cross section. Here $\zeta_{sh} \lesssim 1$ is the fraction of
cold protons inside the SNR shell. The maximum neutrino energy from
pion synchrotron energy losses in the SNR shell magnetic field is
\ba
E_{\nu, sb}^{\rm snr} \approx 7.6\times 10^{9} \left(\frac{\delta_{-1}
v_{9.5} t_d^{3}} {\xi_{B,-2} m_{2.2}} \right)^{1/2}
\Gamma_{300}~\rm{GeV.}
\label{syncoolsnr}
\ea

We calculate the neutrino flux from $pp$ interactions, below the
proton threshold energy at $\Delta$ production, for times $t_d \gtrsim
1$ day as
\ba
E_{\nu}^2 \Phi_{\nu} &=& 1.2 \times 10^{-2}~\frac{\eta_p \zeta_{sh}
f_{pp,0} \kappa L_{\gamma,51} E_{\nu}^2} {D_{27.2}^{2}} \nonumber \\
&& \times \int^{E_{p}^{mx}} dE_p \frac{M_{\nu} (E_p)} {E_p^{2}}~
\rm{GeV~cm}^{-2}~\rm{s}^{-1}; \nonumber \\ M_{\nu}(E_p) &=&
\frac{7}{2} \frac{\Theta \left( \frac{1}{4} \frac{m_{\pi}}{\rm GeV}
\gamma_{\rm cm} \leq \frac{E_{\nu}}{\rm GeV} \leq \frac{1}{4
}\frac{E_p}{\rm GeV} \right)}{ \left( \frac{E_{\nu}}{\rm GeV} \right)
{\rm ln}\left( \frac{10^{11}\, {\rm GeV}}{E_p} \right)};
\label{snr-ppnuflux}
\ea
where $f_{pp,0} \approx 1$ is the neutrino conversion efficiency in
$pp$ interactions and $E_p^{mx} \approx 20 \times E_{\nu,th}^{\rm
sn,m}$ from Eq. (\ref{snr-nuthresh}). The $p\gamma$ neutrino flux
component is similar to the burst neutrino flux and we calculate it as
\ba
E_{\nu}^2 \Phi_{\nu} &=& 3.1 \times 10^{-3} ~\frac{f_{\pi,0}^{\rm snr}
\eta_p \kappa L_{\gamma,51}} {D_{27.3}^{2}} \nonumber \\ && \times
\cases{1; \;\; E_{\nu,sb}^{\rm snr} > E_{\nu} > E_{\nu,th}^{\rm sn,m}
\cr (E_{\nu}/E_{\nu,sb}^{\rm snr})^{-1}; \;\; E_{\nu} >
E_{\nu,sb}^{\rm snr} } \nonumber \\ && \times \;
\rm{GeV~cm}^{-2}~\rm{s}^{-1};
\label{snr-pgnuflux}
\ea 
where $f_{\pi,0}^{\rm snr} \approx 1$ is the fraction of proton energy
lost to pions. We assume that all protons ($\eta_p = 1$) below and
10\% protons ($\eta_p = 0.1$) above the proton threshold energy $20
\times E_{\nu,b}^{\rm burst}$ [Eq. (\ref{burst-nubreak})] escape the
GRB internal shocks in Eqs. (\ref{snr-ppnuflux}) and
(\ref{snr-pgnuflux}) to calculate neutrino flux.

For $t_d \ll 1$ day the calculation needs to include pion energy
losses, due to inverse Compton scatterings in a high density photon
field, before they decay to neutrinos.

\subsection{Precursor}
\label{subsec:precursor}

A precursor neutrino component ($\sim 30$ s duration) is generated
while the GRB jet is still burrowing its way out of the stellar
envelope. We have calculated this component for a similar progenitor
model in detail elsewhere \cite{rmw03b}. There we considered a
presupernova star which had either lost ($R_* \sim 10^{11}$ cm) or
retained its H envelope ($R_* \sim 10^{12}$ cm).  For the simultaneous
SN-GRB scenario, these presupernova models are the same. In the
``supranova'' scenario, the status of the progenitor is uncertain,
since numerical simulations which might provide guidance are not
available so far.  One possibility is that, except for the ejected
envelope, the rest of the star becomes the pulsar plus possibly a disk
prior to collapse to a black hole leading to the GRB, so that a
precursor may not be seen before the GRB event, since there is no
stellar core to burrow through. A second possibility (depending on the
delay) is that, at the time of the GRB, there is some fraction of the
stellar gas which is not part of the ejected shell at radii
intermediate between the shell and the central object, which could
lead to a precursor resembling the above progenitor cases. The gas
column density traversed by the jet is likely to be smaller than in
the simultaneous SN case, reducing the $pp$ contribution and producing
a weaker precursor.

\begin{figure}[htb]
\centerline{\epsfxsize=3.5in \epsfbox{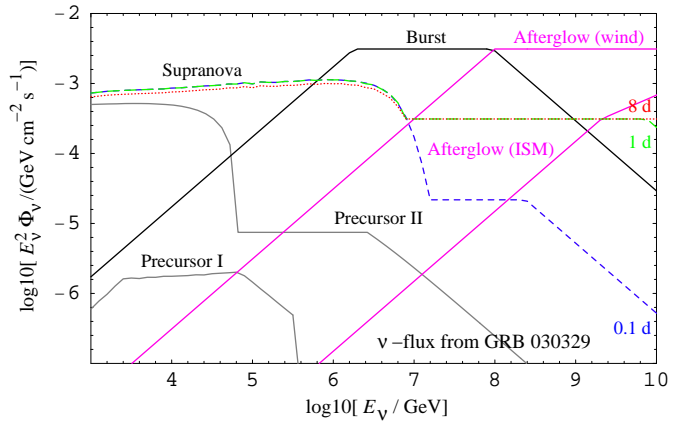}}
\caption{Muon neutrino flux components from GRB 030329 and the
associated SN 2003dh. Possible ``Supranova'' components are plotted
for 0.1 d (short dashed curve), 1 d (long dashed curve) and 8 d
(dotted curve) delays of the SN event prior to the GRB. The
``Precursor'' signals are calculated in the case when the GRB and the
SN events take place simultaneously. The ``Precursor I'' signal is for
a progenitor star which has lost its H envelope.  The ``Precursor II''
signal assumes that the star has retained an H envelope, or the He/C/O
core has expanded, out to a radius $\sim 10^{12}$ cm before the onset
of GRB event. The internal shock ``Burst'' component \cite{wb-burst}
is shown by full dark lines. The ``Afterglow (ISM)'' and ``Afterglow
(wind)'' components, also shown as full dark lines, are calculated for
expansion into a lower density surrounding ISM, and a higher density
wind, respectively. }
\label{fig:nuflux}
\end{figure}

We have plotted muon neutrino flux components at earth from GRB 030329
associated with SN 2003dh in Fig. \ref{fig:nuflux}. The ``Precursor''
components arrive $\sim 30$ s prior to the ``Burst'' component, and
are expected in the case when the GRB and the SN events take place
simultaneously.  The ``Precursor I'' signal assumes the presupernova
stellar radius is $\approx 10^{11}$ cm. The ``Precursor II'' signal is
calculated assuming the star has an envelope, or the He/C/O core has
expanded, out to a radius $\sim 10^{12}$ cm before the onset of the
GRB event.  The ``Afterglow (ISM)'' and ``Afterglow (wind)''
components are calculated when the GRB jet materials run into lower
density ISM and a higher density wind respectively. While the burst
and the afterglow components are based on generic GRB features, the
``Supranova'' components are highly model dependent.  We have plotted
the supranova components assuming that 10\% of the shock accelerated
protons escape internal shocks to interact with the SNR shell which is
up to 8 day old, according to current observational limits. We have
plotted the ``Supranova'' components for 0.1 d (short dashed curve), 1
d (long dashed curve) and 8 d (dotted curve). Note a reduced $p\gamma$
flux contribution ($E_{\nu} \gtrsim 10^7$ GeV) for the 0.1 d case due
to inverse Compton losses of pions. The duration of all components are
$\Delta t \approx 30 \Delta t_{30}$ s.

\section{Detection at earth}
\label{sec:detection}

Neutrino detection at earth requires a large volume of material
because of a small neutrino-nucleon ($\nu N$) cross section
\cite{fmr95, frm96, gandhi96, gandhi98}. Ice Cherenkov detectors such
as AMANDA \cite{amanda}, IceCube \cite{icecube}, RICE \cite{rice} and
ANITA \cite{anita}; water Cherenkov detector such as ANTARES
\cite{antares}; and extended air shower detector such as AUGER
\cite{auger} are poised to detect cosmic neutrinos in different energy
ranges. Several authors have recently calculated expected neutrino
events, based on generic flux models, from individual GRBs
\cite{guetta03}. Here we calculate the number of neutrino events that
one can expect from GRB 030329 and the associated SN 2003dh, based on
flux models described in Sec. \ref{sec:nu-flux}, in an under-ice
Cherenkov detector located at the South Pole as an example.

Neutrinos from a point source in the northern hemisphere travel
through a substantial amount of earth material to reach a detector at
the South Pole. Given a certain zenith angle $\theta$, the column
depth of material is $Z(\theta) = \int_0^L \rho (r;\theta,l) dl$,
where $\rho (r;\theta,l)$ is the density of earth's interior, based on
the Preliminary Reference Earth Model \cite{prem}, at a depth $l$ and
the corresponding earth's interior radius $r$. The upper limit of
integration is the chord length $L = 2R_{\oplus} \cos \theta$ where
$R_{\oplus} \approx 6371$ km is earth's outer radius. The interaction
length for neutrinos inside earth is $\Lambda_{\nu N} ({E_{\nu}}) =
N_A \sigma^{\rm eff}_{\nu N} ({E_{\nu}})$, where $N_A$ is the
Avogadro's number and the effective cross section is given by
\ba
\sigma^{\rm eff}_{\nu N} ({E_{\nu}}) &=& \sigma^{\rm CC}_{\nu N}
({E_{\nu}}) + \sigma^{\rm NC}_{\nu N} ({E_{\nu}}) \nonumber \\ && -
\int_{E_{\nu}}^{\infty} dE'_{\nu} \frac{\Phi'_{\nu}}{\Phi_{\nu}}
\frac{d\sigma^{\rm NC}_{\nu N}}{dE_{\nu}} (E'_{\nu}, E_{\nu});
\label{eff-x} 
\ea
for an incident neutrino flux $\Phi_{\nu}$. Here $\sigma^{\rm CC}_{\nu
N}$ and $\sigma^{\rm NC}_{\nu N}$ are the total charge current (CC)
and neutral current (NC) cross sections respectively and the third
term corresponds to the neutrino regeneration phenomenon due to NC
interactions which degrade a neutrino to lower energies
\cite{berezinskii86, frm96}. For simplicity we have ignored neutrino
regeneration in Eq. (\ref{eff-x}), and the effective cross section is
then the sum of CC and NC cross section $\sigma^{\rm eff}_{\nu N}
\approx \sigma^{\rm CC}_{\nu N} + \sigma^{\rm NC}_{\nu N} =
\sigma^{\rm tot}_{\nu N}$. The survival probability of a neutrino,
from a point source, penetrating through the earth is
\ba
S(E_{\nu}) &=& {\rm exp}~\left[ -Z(\theta)/\Lambda_{\nu N} ({E_{\nu}})
\right] \nonumber \\ &\approx & {\rm exp}~\left[ - N_A \sigma^{\rm
tot}_{\nu N} ({E_{\nu}}) \int_0^L {\rho (r;\theta,l) dl} \right].  
\label{surv-prob}
\ea
We use the total $\nu N$ cross sections (CC and NC) given in
Ref. \cite{gandhi98} to calculate the survival probability.

\begin{figure}[htb]
\centerline{\epsfxsize=3.5in \epsfbox{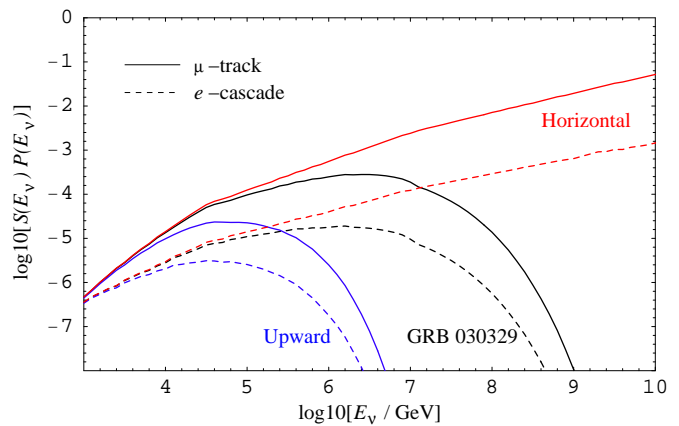}}
\caption{Probability [$S(E_{\nu}) P_{e,\mu}(E_{\nu})$] of neutrino
interaction in the detector volume, after penetrating through earth
from GRB 030329, in a km deep under-ice detector at the South
Pole. Also shown are the probabilities for upward and horizontal
neutrinos for comparison. Probabilities for a muon event and an
electron induced cascade event are shown by solid and dashed curves
respectively.}
\label{fig:probability}
\end{figure}

Neutrinos surviving after penetrating earth must interact with
nucleons inside the effective detection volume to be detected as an
event. Secondary $e$ or $\mu$ produced by the CC interaction
($\nu_{e,\mu} + N \ra e,\mu + X$) then lead to either a high energy
$e$-cascade or $\mu$-track from which a Cherenkov signal can be picked
up in the radio \cite{radio} or optical band \cite{amanda}. The
probability for a $\nu_{e,\mu}$ to interact and produce a secondary
$e,\mu$ of energy $E_{e,\mu}$ above a minimum value $E_{e,\mu}^{\rm
min}$ in the vicinity of the detector is then
\ba 
P_{e,\mu} ({E_{\nu}}) &=& \frac{1}{\sigma^{\rm CC}_{\nu
N}}~\int^{E_{\nu}}_{E_{e,\mu}^{\rm min}} dE_{e,\mu}
~\frac{d\sigma^{\rm CC}_{\nu N}}{dE_{e,\mu}} (E_{\nu}, E_{e,\mu})
\nonumber \\ && \times~\left( 1-{\rm exp} \left[ -N_A \sigma^{\rm
CC}_{\nu N}(E_{\nu}) R_{e,\mu} \right] \right).
\label{int-prob}
\ea
Here $R_e$ is the electron range and is essentially the linear
dimension of the detector. The muon range is given by the analytic
formula
\ba
R_{\mu}(E_{\mu}, E_{\mu}^{\rm min}) = \frac{1}{b} {\rm ln}~
\frac{a+bE_{\mu}}{a+bE_{\mu}^{\rm min}}
\label{muon-range}
\ea
where $a=2 \times 10^{-3}$ GeV cm$^2$ g$^{-1}$ and $b = 4 \times
10^{-6}$ cm$^2$ g$^{-1}$ \cite{dutta01}. We have plotted neutrino
interaction probabilities, in a km deep under-ice detector, after
surviving through earth [$S(E_{\nu}) P_{e,\mu}(E_{\nu})$] from GRB
030329 ($\theta = 68.4^{\circ}$) in Fig. \ref{fig:probability}.
Because of their longer range, the probability for muon tracks (solid
curves) are higher than for electron induced cascades (dashed
curves). We have also plotted the probabilities for horizontal and
upward neutrino events for comparison.

The number of neutrino events, given a flux $\Phi_{\nu} =
d^2N_{\nu}/dE_{\nu} dt$, in a detector of effective area $A_{\rm eff}$
is then
\ba 
N_{e,\mu} = A_{\rm eff} \Delta t \int dE_{\nu}~\frac{d^2N_{\nu}}
{dE_{\nu} dt}~S({E_{\nu}}) P_{e,\mu} ({E_{\nu}}),
\label{event-formula}
\ea
from Eqs. (\ref{surv-prob}) and (\ref{int-prob}). We have used a
duration $\Delta t = 30 \Delta t_{30}$ s for all fluxes. We have
tabulated (see Table \ref{tab:events}) the expected number of muon
tracks and electron cascades from GRB 030329 in an under-ice detector
of $A_{\rm eff} = {\rm km}^2$ for different neutrino flux
components. We used $E_{e,\mu}^{\rm min} = 100$ GeV for these
calculations.

\begin{table}
\caption{\label{tab:events} Neutrino events in a km scale under-ice
detector at the South Pole from the GRB 030329 and the associated SN
2003dh. We have also calculated events assuming similar GRB-SN with
declinations $90^{\circ}$ (upward events denoted by $\uparrow$) and
$0^{\circ}$ (horizontal events denoted by $\ra$) cases. We have not
considered detector response for our calculation.}
\begin{ruledtabular}
\begin{tabular}{l|cc|cc}
Flux & \multicolumn{2}{c|}{TeV-PeV} & \multicolumn{2}{c}{PeV-EeV} \\
Component & $\mu$-track & $e$-cascade & $\mu$ track & $e$-cascade \\
\hline
Precursor I & $9\cdot 10^{-3}$ & $2\cdot 10^{-3}$ & - & - \\ 
& $6\cdot 10^{-3}$ $\uparrow$ & $2\cdot 10^{-3}$ $\uparrow$ & - & - \\
& 0.01 $\ra$ & $2\cdot 10^{-3}$ $\ra$ & - & - \\ \hline
Precursor II & 4.1 & 1.1 & $3\cdot 10^{-3}$ & $2\cdot 10^{-4}$ \\ 
& 2.9 $\uparrow$ & 0.9 $\uparrow$ & - & - \\
& 4.4 $\ra$ & 1.2 $\ra$ & 0.01 $\ra$ & $8\cdot 10^{-4}$ $\ra$ \\ \hline
Burst & 1.8 & 0.2 & 1.4 & 0.1 \\ 
& 0.3 $\uparrow$ & 0.04 $\uparrow$ & - & - \\
& 2.9 $\ra$ & 0.3 $\ra$ & 7.6 $\ra$ & 0.4 $\ra$ \\ \hline
Afterglow & $2\cdot 10^{-4}$  & $2\cdot 10^{-5}$ 
& $2\cdot 10^{-4}$ & $1\cdot 10^{-5}$ \\
(ISM) & $3\cdot 10^{-5}$ $\uparrow$ & $4\cdot 10^{-6}$ $\uparrow$ & - & - \\
& $2\cdot 10^{-4}$ $\ra$ & $2\cdot 10^{-5}$ $\ra$ & 
0.01 $\ra$ & $5\cdot 10^{-4}$ $\ra$ \\ \hline
Afterglow & 0.03 & $3\cdot 10^{-3}$ & 0.05 &  $3\cdot 10^{-3}$ \\
(wind) & $5\cdot 10^{-3}$ $\uparrow$ & $7\cdot 10^{-4}$ $\uparrow$ & - & - \\
& 0.05 $\ra$ & $5\cdot 10^{-3}$ $\ra$ & 1.4 $\ra$ & 0.06 $\ra$ \\ \hline
Supranova & 12.4 & 2.4 & 0.5 & 0.03 \\
0.1 d & 6.1 $\uparrow$ & 1.6 $\uparrow$ & - & - \\
& 14.9 $\ra$ & 2.7 $\ra$ & 1.6 $\ra$ & 0.1 $\ra$ \\ \hline
Supranova & 12.4 & 2.4 & 0.5 & 0.03 \\
1 d & 6.1 $\uparrow$ & 1.6 $\uparrow$ & - & - \\
& 14.9 $\ra$ & 2.7 $\ra$ & 1.9 $\ra$ & 0.1 $\ra$ \\ \hline
Supranova & 10.9 & 2.2 & 0.4 & 0.03 \\
8 d & 5.4 $\uparrow$ & 1.4 $\uparrow$ & - & - \\
& 13.2 $\ra$ & 2.4 $\ra$ & 1.7 $\ra$ & 0.1 $\ra$
\end{tabular}
\end{ruledtabular}
\end{table}

\section{Summary and discussion}
\label{sec:summary}

We have calculated the $\gtrsim$ TeV neutrino signals from gamma-ray
bursts resulting from the collapse of massive stars associated with
supernovae, both in the case where the supernova occurs simultaneously
with, or hours to days in advance of, the $\gamma$-ray event. Longer
SN-GRB time off-sets of weeks to months as considered in the
``supranova" scenario \cite{vie98}, which would have had interesting
consequences for the interpretation of GRB afterglows, are ruled out
by observations of GRB 030329-SN 2003dh \cite{stanek03, matheson03,
SN2003dh}.  Shorter off-sets of days to hours, however, which cannot
be ruled out by optical observations, could be the signature of a
delayed black hole formation from fall-back of gas onto an initial
neutron star.  Such two-step black hole formation could be expected in
intermediate mass ($30\msun \lesssim M_\ast \lesssim 40 \msun$)
stellar progenitor core collapse \cite{macfadyen01}, after delays of
fractions of hours to $\lesssim$ day. Within this range, the time
off-set is so far poorly constrained due to the approximate nature of
the numerical simulations. Thus, any experimental constraints would be
very useful, and neutrino signals such as discussed here would be
essentially the only way to obtain information about the dynamics of
the core collapse. Figure \ref{fig:nuflux} gives the predicted
supranova muon neutrino fluxes for a burst such as GRB 030329 assuming
three different time delays between between the SN and the GRB. For a
simultaneous GRB-SN event, these supranova fluxes would be absent.

The detection of $\gtrsim$ TeV neutrino signals from transient sources
such as GRBs is basically background free \cite{icecube}. The number
of $\mu$-tracks and $e$-cascades from the GRB internal shock [Burst
and Precursor II components] and from interaction of shock-accelerated
GRB protons with the supernova shell (if the later precedes the GRB by
hours to days) are certainly above atmospheric background of $\sim
0.3$ events in the TeV-PeV range. In the PeV-EeV range, $\nu_{\mu}$
events from the Burst component should be detectable over zero
atmospheric background events. We note, furthermore, that the number
of $\nu_{\mu}$ events would be doubled in case there is no flavor
oscillation. The detection of, or upper limits on, such events could
provide important insights both on the collapse dynamics and on the
GRB central engine.

{\it Acknowledgements --} We thank Alexander Heger for helpful
discussion on presupernova star models. This research is supported in
part by NSF AST0098416 and NASA NAG5-13286. EW is partially supported
by a Minerva grant.

\end{document}